%% file: neutrinodm_pev_v2.tex
\documentclass[aps,prl,twocolumn, superscriptaddress,groupedaddress]{revtex4}  
\usepackage{graphicx}  
\usepackage{dcolumn}   
\usepackage{bm}        
\usepackage{amssymb}   

\def\be{\begin{equation}}
\def\ee{\end{equation}}
\newcommand{\bea}{\begin{eqnarray}}
\newcommand{\eea}{\end{eqnarray}}

\newcommand{\pv}{\langle\phi\rangle}
\newcommand{\hv}{\langle H_u^0\rangle}

\begin{document}

\widetext
\leftline{MCTP-14-44}

\title{Neutrino Masses and Sterile Neutrino Dark Matter from the PeV Scale}

\input author_list.tex       

\begin{abstract}
We show that active neutrino masses and a keV-GeV mass sterile neutrino dark matter candidate can result from a modified, low energy seesaw mechanism if right-handed neutrinos are charged under a new symmetry broken by a scalar field vacuum expectation value at the PeV scale. The dark matter relic abundance can be obtained through active-sterile oscillation, freeze-in through the decay of the heavy scalar, or freeze-in via non-renormalizable interactions at high temperatures. The low energy effective theory maps onto the widely studied $\nu$MSM framework.
\end{abstract}

\maketitle

\section{Motivation}
\label{sec:motivation}

A natural resolution of the hierarchy problem has long pointed to the weak scale as the natural scale for supersymmetry. Weak scale supersymmetry was additionally motivated by the WIMP miracle, which offered a natural explanation of dark matter and its observed abundance.
However, the predictions of the most natural setups -- a light Higgs boson, weak scale superpartners (in particular stops and gluinos) within reach of the first run of the LHC, and detection of dark matter at direct detection experiments -- have all failed to materialize, suggesting that the electroweak scale may be fine-tuned after all, and the scale of new physics may lie elsewhere.

Independent of such preconceived notions of naturalness, the measured mass of the Higgs boson at $125$ GeV now provides a direct probe of where this scale might lie. The Higgs mass at one loop with no sfermion mixing in the MSSM is
\be
m_h^2\approx m_Z^2\, \text{cos}^2 2\beta + \frac{3 m_t^4}{4\pi^2v^2}\,\text{ln}(m_{\tilde{t}}^2/m_t^2).
\ee
For tan$\beta\approx\mathcal{O}(1)$, the observed Higgs mass is obtained for sfermion masses at $1-100$ PeV \cite{Giudice:2011cg, ArkaniHamed:2012gw, Arvanitaki:2012ps}. Even prior to the Higgs mass measurement, there were strong arguments for supersymmetry at such high scales from flavor, CP, and unification considerations \cite{Wells:2003tf, ArkaniHamed:2004fb, Giudice:2004tc, Wells:2004di}.

This paper examines whether the neutrino sector and a dark matter candidate can also emerge naturally from the (supersymmetric) PeV scale. Since neutrino masses require physics beyond the Standard Model, a common origin of the Higgs mass, dark matter, and neutrino masses is an extremely attractive prospect.

The traditional explanation of neutrino masses is a seesaw mechanism, involving right-handed, Standard Model (SM)-singlet sterile neutrinos $N_i$ that enable the following terms in the Lagrangian
\be
\label{eq:seesaw}
\mathcal{L}\supset y_{\alpha i} \bar{L}_\alpha H^\dagger_u N_i+M_i \bar{N}^c_i N_i.
\ee
The first term leads to a Dirac mass between the left- and right-handed neutrinos once $H_u$ obtains a vacuum expectation value (vev), and the second term is a Majorana mass for the sterile neutrinos. If $M\gg y \langle H_u\rangle$, the seesaw mechanism gives active neutrino masses at $(y \langle H_u\rangle)^2/M$. GUT scale seesaw models \cite{Minkowski:1977sc, Mohapatra:1980yp, Yanagida:1980xy, GellMann:1980vs, Schechter:1980gr} employ $y\sim\mathcal{O}(1)$ and $M\sim 10^{10}-10^{15}$ GeV, which can explain the small neutrino masses but not dark matter. The low energy counterpart, with all masses below the electroweak scale, has been extensively studied in the effective framework of the Neutrino Minimal Standard Model ($\nu$MSM) \cite{Asaka:2005an, Asaka:2005pn,Asaka:2006ek}, where a keV scale sterile neutrino is a viable warm or cold dark matter candidate (see also \cite{Merle:2013gea}). However, the keV scale is picked by hand, and producing appropriate active neutrino masses requires $y^2\lesssim 10^{-13}$. The purpose of this paper is to explore a modified setup where both active neutrino masses and a dark matter candidate can be realized with predominantly $\mathcal{O}(1)$ couplings and the PeV scale, which is motivated by the Higgs mass measurement as a possible scale for new physics (supersymmetry).

Finally, while not the main motivation of this paper, some recent observational hints add further relevance to this study. A $7$ keV sterile neutrino dark matter candidate can explain the recent observation of a monochromatic line signal at 3.5 keV in the X-ray spectrum of galactic clusters \cite{Bulbul:2014sua}. The observation of neutrinos with PeV scale energies at IceCube \cite{Aartsen:2013bka,Asaka:2005pn} also hint at a possible connection between the neutrino sector and physics at the PeV scale. These can be accommodated in our framework, but are not necessary ingredients, and therefore have been studied in a separate paper \cite{Roland:2015yoa}.


\section{The Model}
\label{sec:model}

As in the $\nu$MSM, the neutrino sector is extended by three right-handed sterile neutrinos $N_i$. Our starting point is the observation that although the $N_i$ are uncharged under the SM gauge group, it is unlikely that they are uncharged under all symmetries of nature (as is traditionally assumed in the seesaw mechanism) if they are to be at the keV-GeV scale, otherwise their masses, unprotected by any symmetry, should naturally be at the Planck or GUT scale. Here we invoke a symmetry to suppress the Majorana masses, but it is worth noting that small Majorana masses can also be technically natural if there are no other sources of lepton number violation. For concreteness, we assume that the $N_i$ are charged under a $U(1)'$, which are ubiquitous in string-inspired models of nature. This immediately forbids the terms in Eq.\,\ref{eq:seesaw}, and the traditional seesaw mechanism does not work. Higher dimensional operators involving the SM and $N_i$ fields can be obtained by coupling the $N_i$ to other fields charged under the $U(1)'$. We introduce an exotic field $\phi$ that carries the opposite charge under $U(1)'$.

As motivated in the previous section, we are interested in a supersymmetric framework, motivated by a possible common origin of the supersymmetry breaking scale and the mass scale that sets the neutrino masses (however, this connection to supersymmetry is by no means necessary). We thus introduce three chiral supermultiplets $\mathcal{N}_i$ for the sterile neutrinos and a chiral supermultiplet $\Phi$, whose spin $(0,~1/2)$ components are labelled $(\tilde{N}_i,N_i)$ and $(\phi, \psi_\phi)$ respectively. With these fields and charge assignments, one is allowed the following higher dimensional operators in the superpotential:
\be
\label{eq:newterms}
W\supset\frac{y}{M_*} L H_u \mathcal{N}\Phi+\frac{x}{M_*}\mathcal{N}\mathcal{N}\Phi\Phi .
\ee
Here $x$ and $y$ are dimensionless $\mathcal{O}(1)$ couplings (neglecting possible flavor structure for now), and $M_*$ is the scale at which this effective theory needs to be UV completed with new physics, such as the scale of grand unification $M_{GUT}$ or the Planck scale $M_P$. Here we have ignored the $(LH_u)^2/M_*$ term, which is of the same dimension, as it is not large enough to produce all active neutrino masses, but we note that it can provide the dominant contribution to the lightest active neutrino mass.

If the scalar $\phi$ obtains a vev at the PeV scale, presumably from the same mechanism that breaks supersymmetry, this breaks the $U(1)'$ and (after $H_u$ also acquires a vev) leads to the following active-sterile Dirac mass and sterile Majorana mass scales
\be
m_D=\frac{y \pv\hv}{M_*},~~~~~m_M=\frac{x \pv^2}{M_*}.
\ee
This results in a modified seesaw mechanism, arising entirely from higher dimensional operators. Below the electroweak scale, the effective theory maps onto the $\nu$MSM with the following sterile and active neutrino mass scales:
\bea
\label{Mas}
m_s &= &m_M=\frac{x \pv^2}{M_*} \label{Ms},\nonumber\\
m_a &=& \frac{m_D^2}{m_M}=\frac{y^2 \hv^2}{x M_*}.
\eea
Note that the two scales are related as
\be
m_s = \frac{1}{m_a} \left(\frac{y \pv \hv}{M_*}\right)^2.
\ee
Fixing the parameters of the theory also determines the mixing angle between the active and sterile sectors:
\be
\label{eq:mixing}
\theta \approx \sqrt{\frac{m_a}{m_s}} = \frac{y \hv}{x \pv}. 
\ee

\begin{figure}
\includegraphics[width=3.3in]{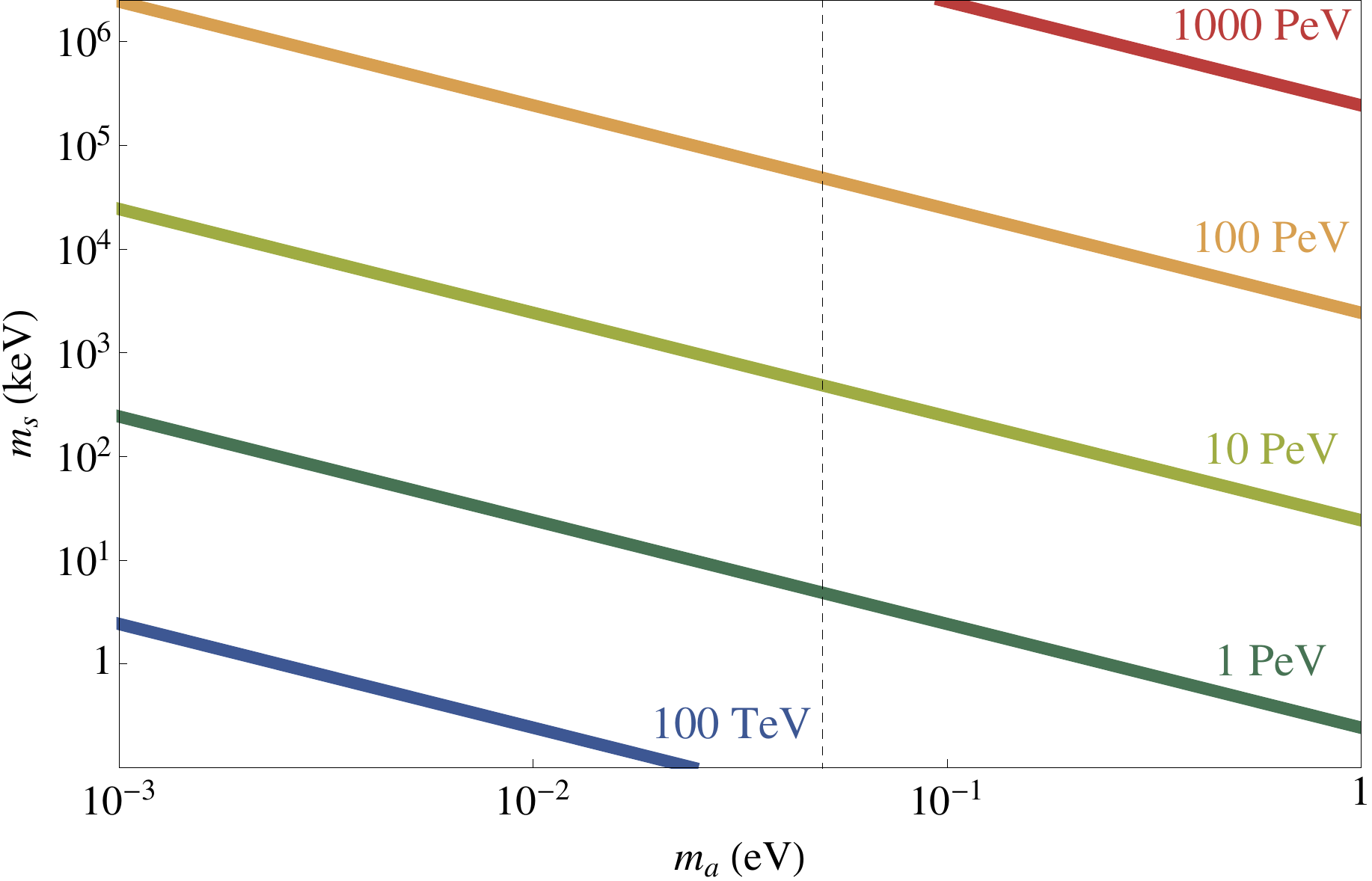}
\caption{\label{fig:scalology} Active and sterile neutrino mass scales for various choices of $y\pv$, with $M_*\,=\,M_{GUT}$, tan$\beta=2 ~(\hv=155.6$ GeV), and $0.001\,\textless\, x \,\textless\, 2$. The dashed vertical line at $m_a=0.05$\,eV is the active neutrino mass scale necessary for consistency with atmospheric oscillation data $\Delta m_{atm}^2=2.3\times10^{-3} \,\text{eV}^2$.}
\end{figure}

Figure\,\ref{fig:scalology} shows possible active-sterile mass scale combinations that result from this framework with $M_*\,=\,M_{GUT}(=\!\!\!10^{16}\, {\rm GeV})$, tan$\beta\!\! =\!\!2 ~(\hv\!\!=\!\!155.6$ GeV), and $0.001\,\textless\, x \,\textless\, 2$ for various values of $y \pv$. This exercise suggests that both an active neutrino mass scale of $\sqrt{2.3\times10^{-3} \,\text{eV}^2}\sim 0.05$ eV, necessary for consistency with atmospheric oscillation data $\Delta m_{atm}^2=2.3\times10^{-3} \,\text{eV}^2$, and a sterile neutrino mass scale of ${\mathcal{O}}$(keV-GeV), necessary for consistency with dark matter and cosmological observations, can emerge naturally in this framework (see \cite{ArkaniHamed:2000bq,ArkaniHamed:2000kj} for similar frameworks that lead to weak scale sterile neutrinos and sneutrino dark matter, see also \cite{Langacker:1998ut}).


\section{Dark Matter and Cosmological Constraints}
\label{sec:cosmology}

Sterile neutrinos are constrained by several cosmological and direct observations, which require  careful treatment. This section provides a brief overview to demonstrate consistency with these constraints and the viability of dark matter; a more extensive and comprehensive study will be presented in a forthcoming paper.

We denote the sterile neutrino dark matter candidate by $N_1$. As $N_1$ couples extremely weakly to the SM fields and is never in thermal equilibrium in the early Universe (we have assumed that possible additional interactions due to the $U(1)'$ are negligible), its relic abundance is not set by thermal freeze-out. Under various conditions, our framework allows multiple production mechanisms for $N_1$.

\textit{Active-sterile mixing: } Production  through active-sterile oscillation at low temperatures, known as the Dodelson-Widrow (DW) mechanism \cite{Dodelson:1993je}, is an inevitable consequence of mixing with the active neutrinos, and is known to produce warm dark matter with relic density approximately \cite{Kusenko:2009up,Dodelson:1993je, Abazajian:2005gj,Dolgov:2000ew, Abazajian:2001nj,Asaka:2006nq}
\be
\Omega_{N_i} \sim0.2\left(\frac{{\text{sin}}^2 \theta}{3\times 10^{-9}}\right)\left(\frac{m_s}{3\,\text{ keV}}\right)^{1.8}.
\ee
Compared to WIMP-motivated cold dark matter (CDM) models, a warm dark matter component might be favorable for a resolution of recent puzzles such as the core vs.\ cusp problem and the ``too big to fail" problem \cite{Lovell:2011rd, BoylanKolchin:2011dk}.

The most stringent constraint on sterile neutrinos comes from X-ray measurements, since a sterile neutrino can decay into an active neutrino and a photon. The decay width for this process is
\be
\Gamma(\nu_s\rightarrow \gamma \nu_a)=\frac{9 \alpha_{EM} G_F^2}{1024\pi^4}\sin^2(2\theta) m_s^5.
\ee
(For listings of various decay channels and widths of keV-GeV scale sterile neutrinos, see e.g. the appendix of \cite{Gorbunov:2007ak}). A combination of X-ray bounds \cite{Boyarsky:2006fg,Boyarsky:2006ag, Boyarsky:2005us,Boyarsky:2007ay,Boyarsky:2007ge} and Lyman-alpha forest data \cite{Seljak:2006qw, Asaka:2006nq, Boyarsky:2008xj} now rule out the prospect of all of dark matter consisting of $N_1$ produced in this manner. However, $N_1$ produced through the DW mechanism can still constitute a significant fraction of the dark matter abundance; an analysis in \cite{Boyarsky:2008xj} showed that $m_s\geq 5$ keV warm component constituting $\leq 60\%$ of the total dark matter abundance is consistent with all existing constraints \cite{Boyarsky:2009ix}. A follow-up study by the same authors \cite{Boyarsky:2008mt} and a more recent study \cite{Harada:2014lma} are also in approximate agreement with these numbers.

\textit{Resonant production:} The presence of a lepton chemical potential in the plasma can lead to resonantly amplified production of $N_1$ \cite{Shi:1998km}, producing a colder non-thermal distribution that can help evade the Lyman-alpha bounds, thereby accounting for all of dark matter. This, however, requires fine-tuning of the order of 1 in $10^{11}$ in the mass difference between the two heavier sterile neutrinos in order to generate the large lepton asymmetry through CP-violating oscillations \cite{Shaposhnikov:2008pf,Roy:2010xq}. Such approximate degeneracy can also provide a mechanism to generate baryon asymmetry in the Universe.

\subsection {Freeze-in Production of Dark Matter}

Beyond these traditional $\nu$MSM approaches, our framework also allows for novel freeze-in processes to contribute to the present abundance of $N_1$ due to the presence of the scalar $\phi$. Since $\phi$ needs to acquire a vev tied to the SUSY breaking scale, it is plausible that it has additional interactions (with the Higgs or supersymmetric sector, for example) that keep it in equilibrium with the thermal bath at high temperatures. In the following, we assume that this is the case (we briefly address the alternate scenario at the end of the section). This opens the following possibilities:

\textit{IR freeze-in:} Once the scalar field obtains a vev $\pv$, the decay channels $\phi\rightarrow N_1\,N_1$ and $H_u\rightarrow N_1 \nu_a$ open up with effective couplings $x_{1}=\frac{2\, x\, \pv}{M_*}$ and $y_1=\frac{y\, \pv}{M_*}$ respectively, resulting in the accumulation of $N_1$ through the freeze-in mechanism \cite{Hall:2009bx,Kusenko:2006rh,Merle:2013wta} until the temperature drops below the mass of the parent particle(s). Assuming $y\,\textless\, x$, the abundance due to  $\phi\rightarrow N_1\,N_1$ is (the abundance due to $H_u\rightarrow N_1 \nu_a$ has a similar form) \cite{Kusenko:2006rh, Petraki:2007gq}
\be
\Omega_{N_1} h^2\sim 0.1 \left(\frac{x_1}{1.4\times10^{-8}}\right)^3 \left(\frac{\pv}{m_{\phi}}\right).
\ee
For $\pv/m_{\phi}\sim\mathcal{O}(1),\,x\sim 1,$ and $\pv\sim 1-100$ PeV, this can be a significant contribution to the dark matter abundance. Indeed, IR freeze-in through decay of heavy particles is a widely used production mechanism for sterile neutrino dark matter \cite{Kusenko:2006rh, Shaposhnikov:2006xi, Petraki:2007gq, Boyanovsky:2008nc,  Merle:2013wta,Frigerio:2014ifa,Kang:2014cia,Shuve:2014doa,Abada:2014zra, Lello:2014yha, Merle:2015oja}.

\medskip

\textit{UV freeze-in:} High temperatures in the early Universe can also overcome the 1$/M_*$ suppression of non-renormalizable interactions from the terms in Eq.\,\ref{eq:newterms}. Dark matter can then be produced through the annihilation processes $\phi\,\phi\rightarrow N_1\,N_1$, $\phi\,H_u\rightarrow \nu_a\,N_1$, $\phi\,\nu_a\rightarrow H_u\,N_1$, and $H_u\,\nu_a\rightarrow \phi\,N_1$. The contribution from $\phi\,\phi\rightarrow N_1\,N_1$ to the dark matter yield and relic density, for instance, are approximately \cite{Elahi:2014fsa, Kusenko:2010ik, Blennow:2013jba, Khalil:2008kp,Dev:2013yza}
\bea
Y_{N_1}&\sim& 5\times 10^{-7} x^2 \left(\frac{T_{RH}\,M_{P}}{M_*^2}\right),\\
\Omega_{N_1} h^2&\sim & 0.1\,x^2\left(\frac{m_s}{\text{GeV}}\right)\left(\frac{1000\,T_{RH}\,M_{P}}{M_*^2}\right).
\eea
If the reheat temperature $T_{RH}$ is sufficiently high, such contributions can also be significant. Such UV freeze-in processes are not present in the $\nu$MSM or its singlet extensions and are novel features of our use of non-renormalizable operators. 

 \begin{figure}
\includegraphics[width=3.4in]{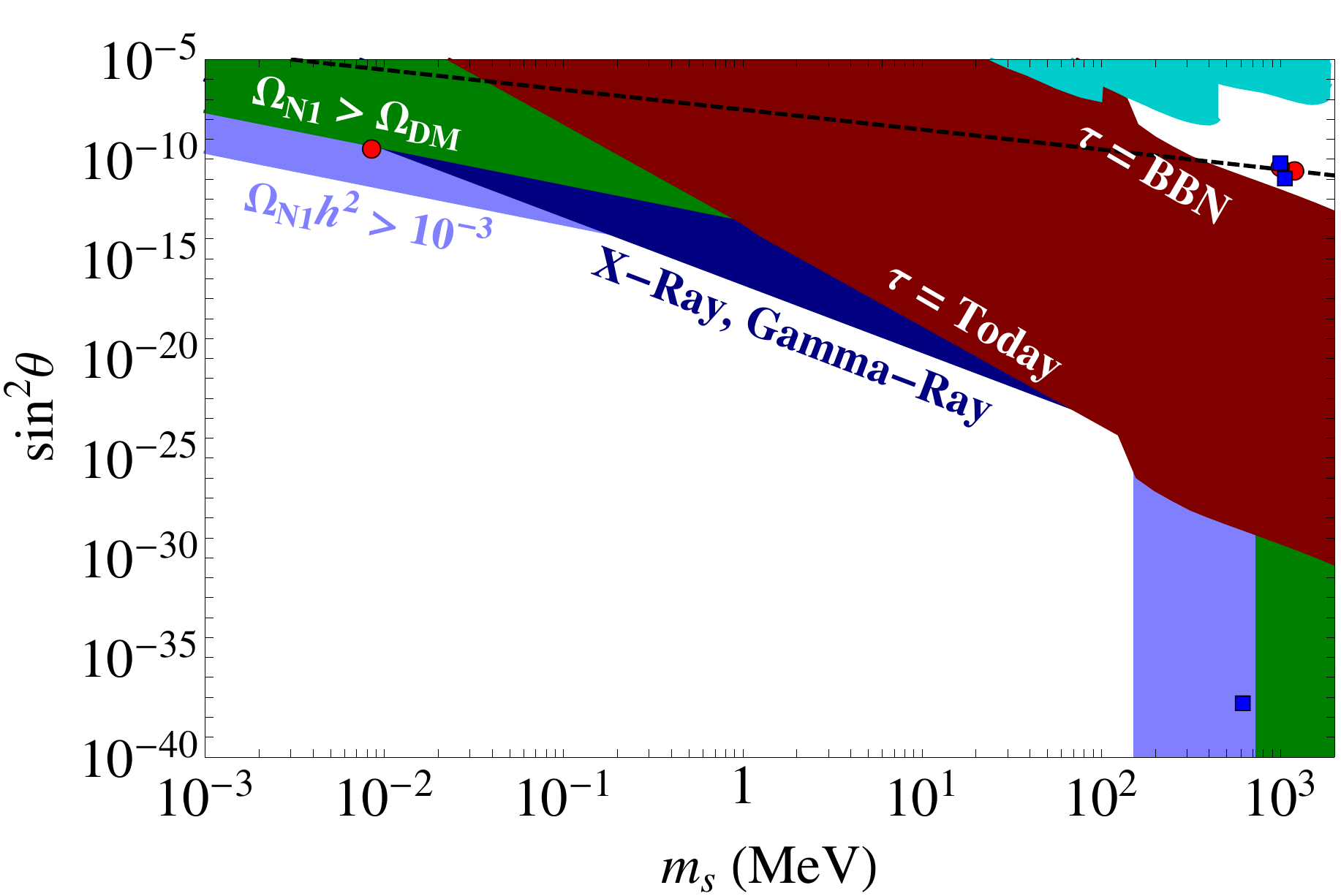}
\caption{\label{fig:constraints} Sterile neutrino parameter space. The black dotted line denotes the combination that yields $m_a\sim 0.05$ eV. In the red region, the lifetime, calculated using several decay channels in \cite{Fuller:2011qy}, is shorter than the age of the Universe but longer than $\tau_{BBN}=1$\,s. Dark matter overcloses the Universe in the dark green region. Dark blue denotes the approximate region ruled out by X-ray and gamma-ray constraints (Fig. 5(right) from \cite{Essig:2013goa}). Cyan regions in top right are constraints from direct searches for heavy neutral leptons (see text for details). The light blue shaded regions denote parameter space where $10^{-3}\leq \Omega h^2\leq 0.12$: the top left region corresponds to DW production, while the bottom right corresponds to IR freeze-in ($M_*\!=\!M_{GUT}\!=\!10^{16}$ GeV and $\pv=m_{\phi}=100$\,PeV everywhere in the plot). Red dots (blue squares) correspond to benchmark scenario A\,(\,B\,) from Table \ref{tab:benchmarks}.}
\end{figure}

We emphasize that the above formulae for IR and UV freeze-in are only approximate, and several $\mathcal{O}(1)$ effects have been ignored. For instance, the dilution of $N_1$ abundance due to entropy production from the decay of other sterile neutrinos \cite{Asaka:2006ek} has not been accounted for.

Figure \ref{fig:constraints} explores the various masses and mixing angles for $N_1$ for which the correct relic density can be obtained (resonant production has been ignored, and $T_{RH}$ is assumed to be sufficiently low that UV freeze-in is negligible). The light blue shaded regions represent parameter space where $10^{-3}\leq \Omega h^2\leq 0.12$; two distinct regions occur, corresponding to two distinct production mechanisms. In the top left region, dark matter is produced through the DW mechanism thanks to significant active sterile mixing sin$^2\theta\sim 10^{-10}$ for $m_s\sim1-10$ keV. In the bottom right region (plotted for $\pv=m_{\phi}=100$\,PeV), $N_1$ is produced via IR freeze-in of $\phi$, where the extremely small mixing angle sin$^2\theta\sim 10^{-38}$ prevents $N_1$ from decaying into SM fields and keeps it safe from gamma-ray constraints \cite{Essig:2013goa}. Other colored regions, described in the figure caption, denote various constraints.

\medskip

\textit{Out of equilibrium $\phi$:} So far, we have relied on a thermal abundance of $\phi$ from which to produce dark matter, which we took to be a plausible scenario. However, even when $\phi$ has no additional interactions that keep it in equilibrium with the thermal bath, so that there is no initial abundance of $\phi$, freeze-in can still provide the desired dark matter abundance. 

This can occur provided an abundance of $\phi$ gradually builds up from the annihilation process $H_u\,\nu_i\rightarrow \phi\,N_j$ if the temperature in the early Universe is sufficiently high to overcome the $1/M_*$ suppression. Note that this process cannot directly produce a large abundance of the dark matter candidate $N_1$ since the $y_{i1}$ couplings corresponding to $H_u\,\nu_i\rightarrow \phi\,N_1$ are required to be extremely small in order to prevent large mixing between $N_1$ and the active neutrinos (which would make $N_1$ short-lived). However,  $y_{i2},y_{i3}$ can be $\mathcal{O}$(1), so an abundance of $\phi$ can be built up. The crucial difference here compared to the equilibrium case is that, given the absence of significant couplings to the particles in the thermal bath, $\phi$ decays dominantly into sterile neutrinos. Hence the entire $\phi$ abundance is converted into sterile neutrinos, with branching fractions proportional to the sterile neutrino masses. The relic abundance of $N_1$ that results from this process is estimated to be 
\be
\Omega_{N_1} h^2\sim 0.1 \sum_{i,j} y_{ij}^2 \left(\frac{m_s}{\text{GeV}}\right)\left(\frac{1000\,T_{RH}\,M_{P}}{M_*^2}\right) Br(\phi\rightarrow N_1 N_1)
\ee

Hence, with a high enough reheat temperature $T_{RH}$, one can obtain the correct dark matter abundance even when $\phi$ does not have any significant additional interactions. 

Finally, we note parenthetically that since the connection to the PeV scale was inspired by considerations of a supersymmetric sector, a stable or sufficiently long-lived superpartner can also account for an $\mathcal{O}(1)$ (cold) fraction of dark matter, as could axions.

\subsection {Other Constraints}

The neutrino sector of the theory also contains two other sterile neutrinos $N_2$ and $N_3$. As in the $\nu$MSM, these mix with the two heavier active neutrinos to produce their masses. In contrast, the dark matter candidate $N_1$ cannot fully participate in the seesaw as various constraints (see Fig. \ref{fig:constraints}) force a suppression of its mixing with the active neutrinos, leaving the lightest active neutrino essentially massless. These generic features of the $\nu$MSM are also present in our framework. The decays of $N_2, N_3$ are constrained by several recombination era observables \cite{Kusenko:2009up, Hernandez:2014fha, Vincent:2014rja,Vincent:2014rja}, hence they are generally required to decay before Big Bang Nucleosynthesis (BBN), which forces $\tau_{N2,N3}\lesssim 1$s and consequently $m_{N2,N3}\gtrsim\mathcal{O}(100)$ MeV. Several direct searches for heavy neutral leptons with significant mixing with the SM also place bounds on their lifetimes. These experiments look for sterile neutrino production in the decay of charged mesons by detecting additional peaks in the charged lepton spectrum or the charged decay products of the sterile neutrinos \cite{PIENU:2011aa,Bergsma:1985is,Ruchayskiy:2011aa,Bernardi:1985ny,Bernardi:1987ek,Vaitaitis:1999wq}. These BBN and direct search constrained regions are shown in Figure \ref{fig:constraints} as red and cyan regions respectively. For the direct search bounds, the two bumps on the left are derived from results from the PS191 experiment \cite{Ruchayskiy:2011aa}, while the third bump is derived from results from the NuTeV experiment \cite{Vaitaitis:1999wq}, and we have simply replotted the bounds on mixing angles from plots in the corresponding papers.

The final ingredient in the theory is the scalar $\phi$. In the early Universe, its annihilation and decay can contribute to a frozen-in abundance of $N_1$, as discussed earlier. Its present day interactions are all suppressed by the high scale $M_*$ and should therefore be too small to probe experimentally, although production in high energy astrophysical processes could lead to rare but possibly observable signatures.

\section{Benchmark Scenarios}

As proof of principle, this section presents two benchmark scenarios in our framework that produce active neutrino masses as well as a sterile neutrino dark matter candidate. We have used the Casas-Ibarra parameterization \cite{Casas:2001sr} with a normal hierarchy of active neutrino masses to verify that the measured mass differences and mixing angles of the PMNS matrix can be reproduced.

\begin{table*}[ht]
\begin{ruledtabular}
\begin{tabular}{ccccccc}
Benchmark &$\pv$ & $\bf{Y}$ & diag\,(\,$\bf{X}$\,) & $m_a$ (eV) & $m_s$ & $\Omega_{s} h^2$~ \\ \hline \\
 A & 79.4 PeV & $\left(
\begin{array}{ccc}
-1.70 & -0.20 & 9\times10^{-5}\\
1.49 & -3.96 & -3\times10^{-5}\\
3.91 & -2.21 & 5\times10^{-5}\\
\end{array}
\right)$ & $\begin{array}{c} 1.91\\1.58\\0.000013\\ \end{array}$ & $\begin{array}{c}0.049\\0.0087\\2.4\times 10^{-6}\end{array}$ & $\begin{array}{c} 1.2 ~\text{GeV} \\ 1.0~\text{GeV} \\ 8.5 ~\text{keV} \\ \end{array}$ & 0.058~ \\
\\
 B& 85.1 PeV & $\left(
\begin{array}{ccc}
-1.31 & 0.73 & \sim 0\\
-1.25 & -3.71 & \sim 0\\
1.45 & -3.65 & \sim 0\\
\end{array}
\right)$ & $\begin{array}{c}1.46\\1.38\\0.85\end{array}$ &  $\begin{array}{c}0.049\\0.0087\\\sim 0\end{array}$ & $\begin{array}{c} 1.1~\text{GeV} \\ 1.0~\text{GeV} \\ 617~ \text{MeV}\\ \end{array}$ & 0.11~ \\
\\
\end{tabular}
\end{ruledtabular}
\caption{\label{tab:benchmarks}The two benchmark scenarios. Both use $M_*=M_{GUT}=10^{16}$ GeV and tan$\beta=2$, corresponding to $\hv=155.63$ GeV. Benchmark A contains a keV scale warm dark matter candidate produced through the DW mechanism. Benchmark B consists of a GeV scale candidate produced through freeze-in from $\phi$ decay, which can be in or out of equilibrium with the thermal bath (see text).}
\end{table*}

Restoring the full flavor structure, the neutrino mass matrix is a $6\times 6$ entity, with $x$ and $y$ in Eq.\,\ref{eq:newterms} now promoted to $3\times3$ matrices $\bf{X}$ and $\bf{Y}$. The neutrino mass matrix reads
\be
M_{\nu}=
\left(
\begin{array}{cc}
0 & \frac{\pv\hv}{M_*} \bf{Y}\\
\frac{\pv\hv}{M_*} \bf{Y}^{\dagger} & \frac{\pv^2}{M_*} \bf{X}\\
\end{array}
\right).
\ee
The $N_i$ basis can be chosen such that $\bf{X}$ is diagonal.

The two benchmark scenarios are listed in Table \ref{tab:benchmarks}. Both use $M_*=M_{GUT}=10^{16}$ GeV and tan$\beta=2$, corresponding to $\hv=155.63$ GeV.

\subsection{Benchmark A : DW Production}

This scenario has a warm dark matter candidate with mass $8.5$ keV, with DW production giving $53\%$ of the observed dark matter abundance. Note that since $x\approx10^{-5}$, both IR and UV freeze-in are ineffective, but a particle from the supersymmetric sector or the axion could account for the remaining dark matter abundance. The two heavier steriles are at 1 GeV and decay before BBN; the three steriles are plotted as red dots in Figure \ref{fig:constraints}. The hierarchy of five orders of magnitude in the entries of $\bf{X}$ is necessitated by the hierarchy between the keV mass of the dark matter candidate and the GeV scale mass of the heavier steriles, which need to be heavy enough to decay before BBN. The entries of $\bf{Y}$ contain a similar hierarchy to ensure that the dark matter candidate does not mix excessively with the active sector. While a coupling of $\mathcal{O}(10^{-5})$ appears unnatural, such a small coupling already appears in nature in the form of the electron Yukawa, and is therefore perhaps not unrealistic. The lightest active neutrino is essentially massless, as is characteristic in the $\nu$MSM with a keV scale sterile neutrino dark matter candidate. 

\subsection{Benchmark B: Freeze-in Production}

This scenario allows for the scalar $\phi$ to be either in or out of equilibrium with the thermal bath; we consider both cases, and the parameters listed in Table \ref{tab:benchmarks} apply to both. If $\phi$ has additional interactions that keep it in equilibrium with the thermal bath in the early Universe, the dark matter relic density is insensitive to the temperature of the early Universe (as long as it is high enough to produce $\phi$) and is achieved through (IR) freeze-in. Otherwise, an abundance of $\phi$ has to be built up from UV freeze-in as discussed in the previous section, which requires a high reheat temperature, and its decays produce an abundance of $N_1$. In this case, once the remaining parameters are specified, the temperature $T_{RH}$ can be appropriately chosen to yield the correct abundance of dark matter. For the parameters listed for Benchmark B in Table \ref{tab:benchmarks}, the temperature required is $T_{RH}\approx10^{9}$ GeV.

In contrast to Benchmark A, all entries in $\bf X$ are $\mathcal{O}$(1), and all sterile neutrinos have $\sim 1$ GeV mass (represented by blue squares in Figure \ref{fig:constraints}).  In order to make the dark matter candidate sufficiently long-lived and evade gamma-ray constraints \cite{Essig:2013goa}, its mixing with the active neutrinos must be suppressed to essentially zero, reflected in the third column of $\bf Y$. While this appears unnatural, note that it is admissible to set these numbers to exactly zero, hence this structure could be invoked due to an underlying symmetry, rendering it technically natural. Such considerations are only necessary if we insist on promoting $N_1$ to a long-lived dark matter candidate; otherwise, ${\cal O}(1)$ couplings are allowed.

\subsection{Summary}

In summary, this paper has presented a new framework that constitutes a realistic description of active neutrino masses and keV-GeV scale sterile neutrino dark matter emerging naturally from new physics at the PeV scale, which maps on to the widely studied $\nu$MSM at low energies. A more extensive study of the details of this framework, including dark matter, cosmological aspects, and observable signatures, will be presented in forthcoming work.

\medskip
\textit{Acknowledgements: }The authors are supported in part by the DoE under grants DE-SC0007859 and DE-SC0011719.
\bibliography{neutrino_bibliography}

\end{document}

%% file: author_list.tex
\affiliation{Michigan Center for Theoretical Physics, University of Michigan, Ann Arbor MI 48109, USA}
\author{Samuel B. Roland}
\author{Bibhushan~Shakya}
\author{James D. Wells} 


\vskip 0.25cm